\documentclass{elsart}

\usepackage{epsfig}
\usepackage{amssymb}

\begin{document}

\begin{frontmatter}

  \title{A study of the diffusion pattern in $N=4$ SYM at high energies}

\author[Madrid]{F. Caporale},
\author[Valencia]{G. Chachamis},
\author[Madrid]{J. D. Madrigal},
\author[Cosenza]{B. Murdaca},
\author[Madrid]{A. Sabio Vera}

\address[Madrid]{Instituto de F{\'\i}sica Te{\'o}rica UAM/CSIC \& \\Universidad Aut{\'o}noma de Madrid, E-28049 Madrid, Spain}
\address[Valencia]{Instituto de F{\'\i}sica Corpuscular UVEG/CSIC, 46980 Paterna (Valencia), Spain.}
\address[Cosenza]{Dipartimento di Fisica, Universit{\`a} della Calabria \& \\Istituto Nazionale di Fisica Nucleare, Gruppo collegato di Cosenza,\\ I-87036 Arcavacata di Rende, Cosenza, Italy}

\begin{abstract}
In the context of evolution equations and scattering amplitudes in the high energy limit of 
the $N=4$ super Yang-Mills theory we investigate in some detail the BFKL gluon Green 
function at next-to-leading order. In particular, we study its collinear behaviour in terms of 
an expansion in different angular components. We also perform a Monte Carlo simulation 
of the different final states contributing to such a Green function and construct the 
diffusion pattern into infrared and ultraviolet modes and multiplicity distributions, making emphasis in separating the gluon contributions from those of scalars and gluinos. We find 
that the combined role of the non-gluonic degrees of freedom is to improve the collinear behavior and reduce the diffusion into ultraviolet regions while not having any effect on the average multiplicities or diffusion into the infrared. In terms of growth with energy, the non-zero conformal spin components are mainly driven by the 
gluon terms in the BFKL kernel. For zero conformal spin (Pomeron) the effect of the scalar 
and gluino sectors is to dramatically push the Green function towards higher values. 
\end{abstract}

\end{frontmatter}

In this brief letter we address the question of what is the effect in the BFKL equation~\cite{BFKL1}   for 
a supersymmetric theory ($N=4$ super Yang-Mills (SYM)) of the non-gluonic contributions 
to the kernel. We perform the analysis at next-to-leading order (NLO), where the QCD~\cite{Fadin:1998py} and $N=4$ 
SYM evolution equations start being different~\cite{joseagusgreg}. This is a topical subject~\cite{several} 
given the recent 
works by Costa, Goncalves and Penedones~\cite{Costa:2012cb} on conformal Regge theory, and by Kotikov and Lipatov~\cite{LevTolya} on the structure of the BFKL Pomeron at strong coupling in the 
same theory. The available kernel has been calculated at NLO but 
it is feasible to construct higher order corrections to it in the near future, in the planar limit of color space. It will then be interesting to extend our calculations to those higher order equations  with the target to understand how the diffusion features of the BFKL equation change from a resummation at weak coupling to the strong coupling results of Polchinski-Strassler~\cite{Polchinski:2002jw}, Costa-Goncalves-Penedones~\cite{Costa:2012cb} and Kotikov-Lipatov~\cite{LevTolya}. 

Without further introduction let us set our notations. It is well-known that the BFKL gluon 
Green function (the solution to the BFKL equation at NLO for forward scattering) can be 
written as an expansion on azimuthal angle Fourier components, {\it i.e}
\begin{eqnarray}
f\left(\vec{k}_a,\vec{k}_b, {\rm Y}\right)&=&
\sum_{n=-\infty}^{\infty} f_n\left(|\vec{k}_a|,|\vec{k}_b|, {\rm Y}\right) e^{i n \theta}.
\label{expansion2}
\end{eqnarray}
We consider the amplitude for off-shell reggeized gluons with transverse momenta 
$\vec{k}_{a,b}$ with a relative azimuthal angle $\theta$ and a separation in rapidity space 
given by $Y$. Evolution with energy of this gluon Green function corresponds to evolution in
 this variable $Y$. 

The coefficients of the $\theta$-expansion can be written in the form
\begin{eqnarray}
f_n\left(|\vec{k}_a|,|\vec{k}_b|, {\rm Y}\right) &=& \int_0^{2 \pi} \frac{d\theta}{2 \pi} \, 
f\left(\vec{k}_a,\vec{k}_b, {\rm Y}\right) \cos{\left(n \theta\right)} \nonumber\\
&=& \frac{1}{\pi |\vec{k}_a| |\vec{k}_b|} 
\int \frac{d \gamma}{2 \pi i} 
\left(\frac{\vec{k}_a^2}{\vec{k}_b^2}\right)^{\gamma-\frac{1}{2}}
e^{\omega_n (a,\gamma) {\rm Y}},
\end{eqnarray}
where we have performed a Mellin transform in transverse momentum space, with the 
following eigenvalue for the BFKL kernel~\cite{LevTolya}:
\begin{eqnarray}
\omega_n (a,\gamma) &=& 
\xi_\Theta \, \left( 2 \psi(1) - \psi\left(\gamma+\frac{n}{2}\right)-\psi\left(1-\gamma+\frac{n}{2}\right)\right) 
+ a^2 \, \frac{3}{2} \, \zeta (3)  \\
&&\hspace{-1cm}+\frac{a^2}{4} \left[\psi''\left(\gamma+\frac{n}{2}\right)+
\psi''\left(1-\gamma+\frac{n}{2}\right)-2 \Phi(n,\gamma)-2 \Phi(n,1-\gamma)\right] \nonumber \\
&&\hspace{-2cm}-\frac{\Theta \, a^2 \, \pi^2 \cos{\left(\pi \gamma\right)}}{4 \sin^2\left(\pi \gamma\right)\left(1-2\gamma\right)}\left\{\left(3+\frac{2+3\gamma\left(1-\gamma\right)}{\left(3-2\gamma\right)\left(1+2\gamma\right)}\right)\delta_n^0-\frac{\gamma\left(1-\gamma\right)\delta_n^2 }{2\left(3-2\gamma\right)\left(1+2\gamma\right)}\right\},\nonumber
\label{eigenvaluesf}
\end{eqnarray}
with 
\begin{eqnarray}
\Phi(n,\gamma) &=& \sum_{k=0}^{\infty} \frac{(-1)^{(k+1)}}{k+\gamma+\frac{n}{2}}
\left(\frac{}{}\psi'(k+n+1)-\psi'(k+1)\right.\\
&&\hspace{-1cm}+\left.(-1)^{(k+1)} \left(\beta'(k+n+1)+\beta'(k+1)\right)
-\frac{\left(\psi(k+n+1)-\psi(k+1)\right)}{k+\gamma+\frac{n}{2}}\right),\nonumber\\
4 \beta'(z) &=& \psi'\left(\frac{1+z}{2}\right)-\psi'\left(\frac{z}{2}\right).
\end{eqnarray}

The function of the coupling 
\begin{eqnarray}
\xi_\Theta &=& a +  \frac{a^2}{4}\left(\frac{1}{3}-\frac{\pi^2}{3}+\Theta\right),
\label{xi}
\end{eqnarray}
has been introduced where $\Theta=1$ corresponds to diagrams with only gluons and 
$\Theta=0$ to the full $N=4$ SYM result (there are cancellations due to the gluino and scalar contributions). We have not separated the scalars from the gluinos for simplicity since the 
expressions for the kernel, especially in $k_t$ space, are rather complicated and do not add much 
information to our results. For this first, analytic, study we have not considered the contributions to the 
running of the coupling in the gluonic kernel since we wanted to work with true eigenfunctions also 
at NLO and keep all the terms in the kernel diagonal in $\gamma$ space. But when working with a 
Monte Carlo code in the second part of our analysis we have included these running coupling 
terms (note that there are running contributions both in the gluon and gluino/scalars sectors independently, which cancel each other in the complete $N=4$ SYM kernel).

\begin{figure}[tbp]
  \centering
  \epsfig{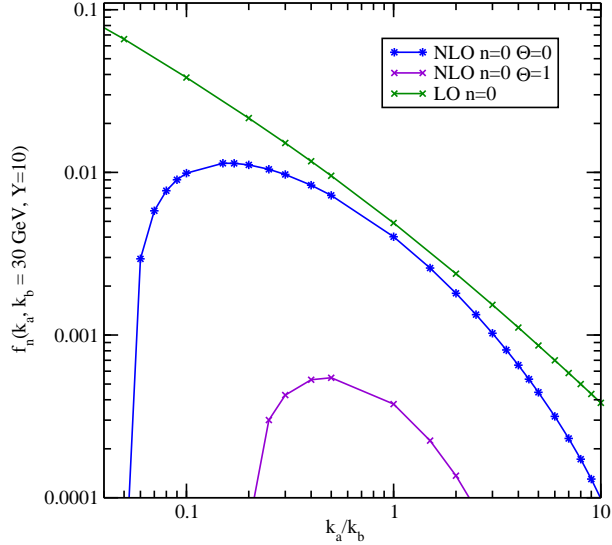}
  \caption{Collinear behaviour of the gluon Green function for $n=0$.}
  \label{Coll_n0}
\end{figure}
Let us first scan the (anti)-collinear regions where one of the virtualities of the external reggeized gluons 
is much larger than the other. This is parameterized by the ratio $k_a/k_b$ being away from one in the 
Green function values plotted in Fig.~\ref{Coll_n0} (we have fixed the coupling to 0.2, $Y=10$ and 
$k_b=30$ GeV, but the features here discussed are generic). 
We have first focussed on the $n=0$ component (azimuthal angle averaged kernel),  which 
corresponds to Pomeron exchange and is the relevant one when going to the strong coupling 
limit~\cite{Brower:2006ea}. 
We observe that the LO Green function has good collinear behavior while the NLO lines go rapidly to zero 
when $k_a$ is very different from $k_b$. 
\begin{figure}[tbp]
  \centering
  \epsfig{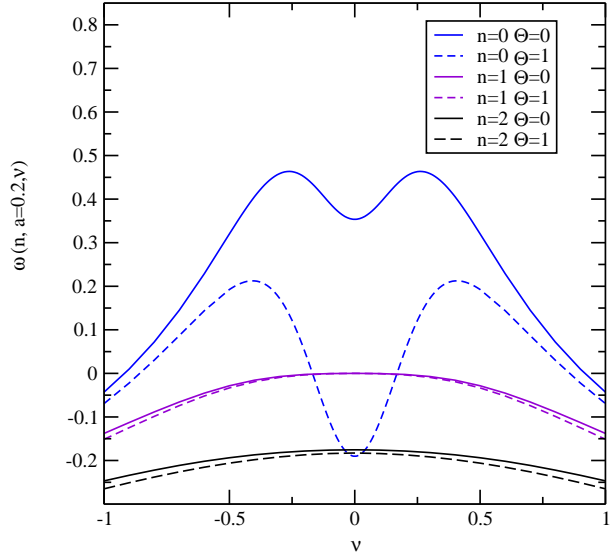}
  \caption{Eigenvalue of the BFKL kernel for different conformal spins, $n$.}
  \label{Omega_n}
\end{figure}
This is a manifestation of the double maxima present in the eigenvalue of the kernel, as it can be seen in Fig.~\ref{Omega_n} where we have set $\gamma= 1/2 + i \nu$~\cite{doublemaxima}. Since the eigenvalue is smaller, even negative at $\nu=0$, for the purely gluonic case it is in this case that the collinear behavior is worse (in the 
sense that the Green function hits negative values for values of $k_a/k_b$ closer to one). The 
effect of the scalar and gluino pieces is to greatly improve the convergence, making the NLO corrections 
to the LO result not too large in a wide range of virtuality space. 

\begin{figure}[tbp]
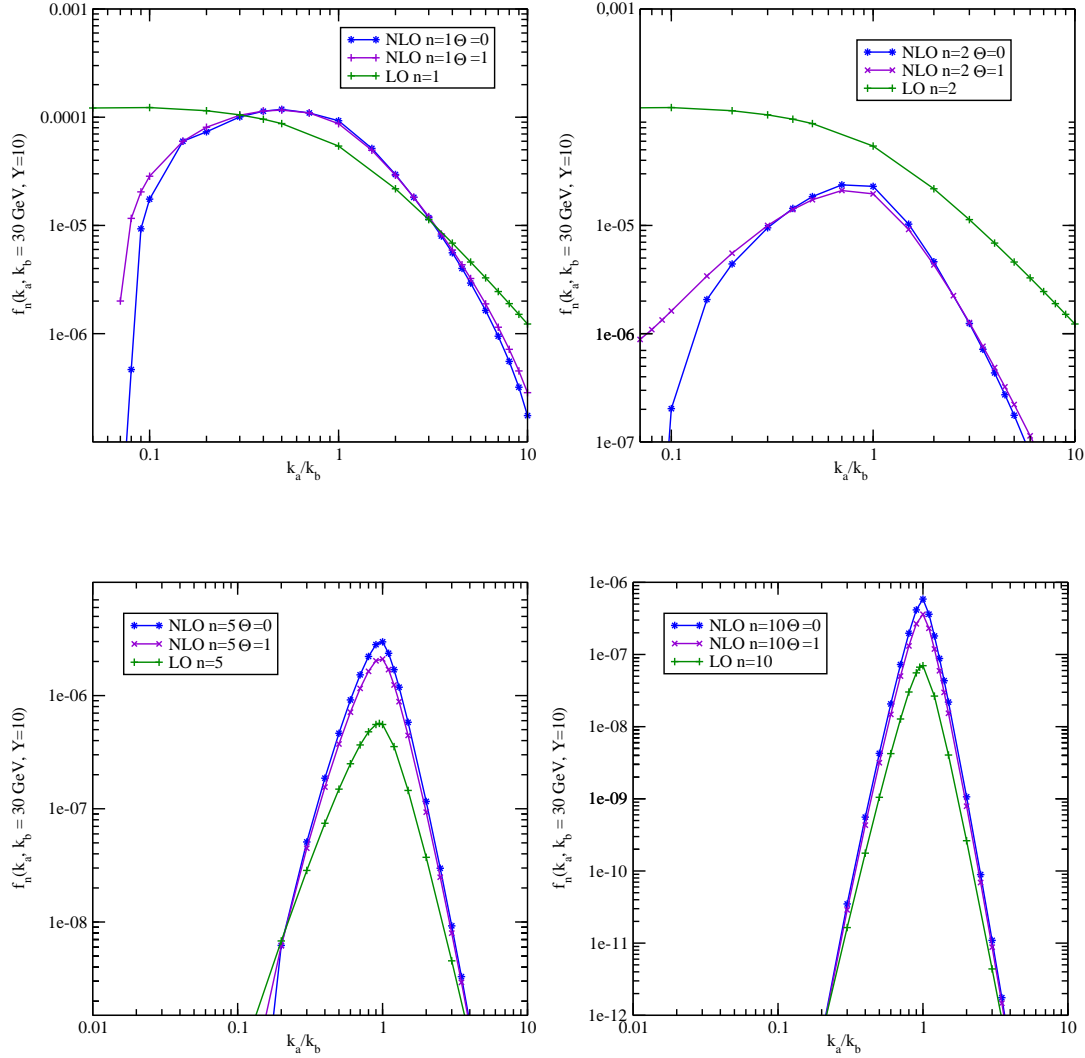

  \centering
  \epsfig{width=7cm,file=Coll_n1}$~~~$\epsfig{width=7cm,file=Coll_n2}\\
  \vspace{1.2cm}
  \epsfig{width=7cm,file=Coll_n5}$~~$\epsfig{width=7cm,file=Coll_n10}
  \caption{Collinear behaviour of the gluon Green function for different values of $n$.}
  \label{Coll_n}
\end{figure}
In Fig.~\ref{Omega_n} we also plot the $n=1,2$ eigenvalues and notice that the effect of the gluino and 
scalar contributions is very small. It is then natural to observe that, at the level of the gluon Green function,  
in Fig.~\ref{Coll_n} the plots with $\Theta=0,1$ are very similar for all $n > 0$. 
\begin{figure}[tbp]
\centering
\epsfig{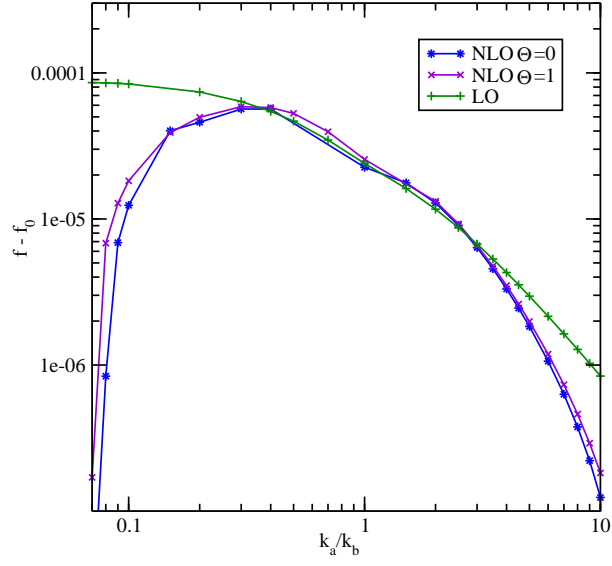}
\caption{Collinear behaviour of the gluon Green function subtracting the $n=0$ component.}
\label{f-f0}
\end{figure}
We find it instructive to plot 
the full gluon Green function with all $n$ components but subtracting the  $n=0$ term while 
showing the collinear shape in Fig.~\ref{f-f0}. The perturbative convergence of the BFKL expansion for the $n \neq 0$ contributions is very good since the NLO corrections are very small in a very wide range of the plot. 
\begin{figure}[tbp]
  \centering
\vspace{2cm}
  \epsfig{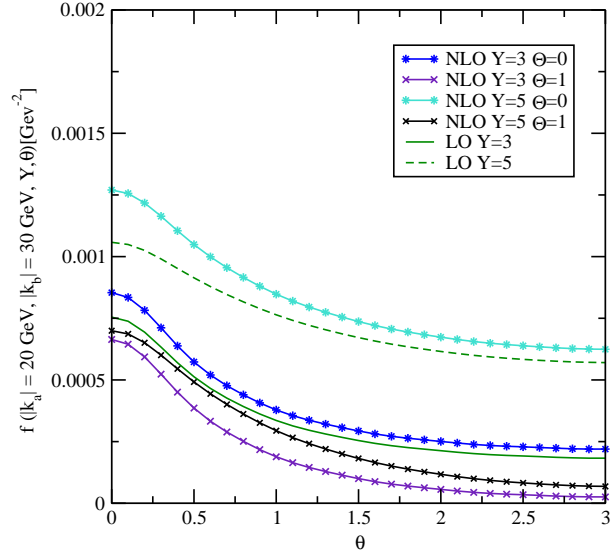}
  \caption{The gluon Green function versus $\theta$.}
  \label{f_vs_angle_tot}
\end{figure}
We collect the full angular information in a single plot in Fig.~\ref{f_vs_angle_tot}. We have fixed 
$k_a=20$ GeV, $k_b=30$ GeV and $Y=3, 5$. The full NLO SUSY results are very similar in shape to the 
LO lines. An intriguing feature is that the Green function for the gluon contributions reaches a much smaller value at $\theta=\pi$ than when scalars and gluinos are included in the analysis ($\partial f / \partial \theta$ is much more negative in the $N=4$ SYM case for $\theta > 1$).

\begin{figure}[tbp]
  \centering
  \epsfig{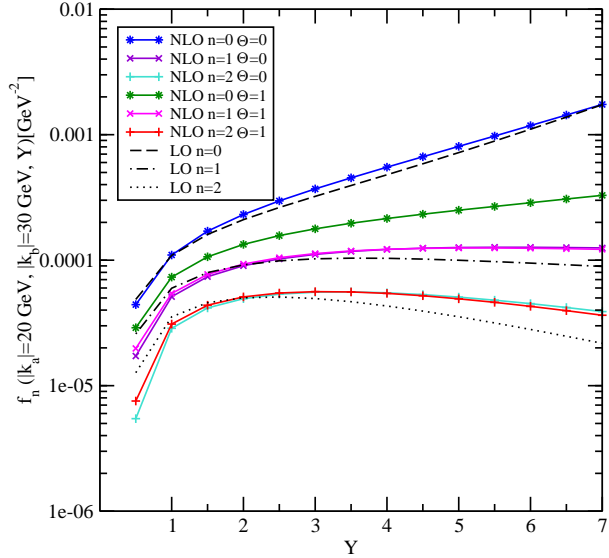}
  \caption{Growth with energy of the gluon Green function for different conformal spins $n$.}
  \label{fn_vs_Y_Confr}
\end{figure}
To conclude this part of the analysis we can investigate the growth with energy of the gluon Green function for different conformal spins $n$. This is done in Fig.~\ref{fn_vs_Y_Confr} where we can see that the LO and the full NLO SUSY results are surprisingly similar for the range of $Y$ we have chosen to plot (for $n=0$). It is clear that the scalar and gluino contributions do push the Green function to higher values. This implies that they generate 
a larger amount of real emission and/or reduce the relative weight of the virtual diagrams, mainly via their contribution  
to the gluon Regge trajectory. For the coefficients associated to $n>0$ the non-gluonic terms do not modify the 
gluonic ones, they give a very small contribution. As in QCD, only the $n=0$ component, associated to the 
hard Pomeron, grows with energy. 

Let us highlight a very interesting property of the NLO eigenvalue, already pointed out 
in Ref.~\cite{Vera:2006un} (Ec.~(38)) for the QCD case. When $n=1$ the asymptotic intercept 
(at $\gamma = 1/2$) can be written as
\begin{eqnarray}
\omega_{n=1} \left(a, \gamma=\frac{1}{2}\right) &=& a^2 \left(\frac{3}{2} \zeta (3) + \frac{1}{2} \psi'' (1) 
- \Phi \left(1,\frac{1}{2} \right)\right) ~=~0,
\end{eqnarray}
which is equal to zero, independently of the scalar and gluino terms, {\it i.e.}, it is an effect only associated to the 
gluon sector. A similar feature was found in QCD, where the quark contributions to this intercept were always 
multiplying the LO eigenvalue, which is also zero for $(\gamma,n)=(1/2,1)$. The fact that this intercept is zero at LO and 
NLO seems to indicate that it is protected by some symmetry not broken by radiative 
corrections. It would be instructive to 
find out if it is present in the strong coupling limit and its connection to all-orders corrections to Odderon exchange in QCD and SUSY theories (see Ref.~\cite{Kovchegov:2012rz} for a related discussion).

In order to proceed further and obtain more exclusive information from the different pieces of the SUSY 
NLO BFKL kernel now we use a different, more numerical, method (see Ref.~\cite{Andersen:2004uj} for a related work). We Mellin transform in rapidity space, 
{\it i.e.}
\begin{equation}
\label{Mellin}
f \left({\bf k}_a,{\bf k}_b, Y\right) 
= \frac{1}{2 \pi i}
\int_{a-i \infty}^{a+i \infty} d\omega ~ e^{\omega Y} f_{\omega} 
\left({\bf k}_a ,{\bf k}_b\right).
\end{equation}
We can then write the NLO BFKL equation in momentum representation:
\begin{eqnarray}
\label{nll}
\left(\omega - \omega \left({\bf k}_a^2,\lambda^2\right)\right) f_\omega \left({\bf k}_a,{\bf k}_b\right) &=& \delta^{(2)} \left({\bf k}_a-{\bf k}_b\right)\\
&&\hspace{-4cm}+ \int d^2 {\bf k} \left(\frac{ \xi_\Theta }{\pi {\bf k}^2}\theta\left({\bf k}^2-\lambda^2\right)+
{\mathcal{K}} \left({\bf k}_a,{\bf k}_a+{\bf k}\right)\right)f_\omega \left({\bf k}_a+{\bf k},{\bf k}_b\right),\nonumber
\end{eqnarray}
where $\lambda$ is a mass parameter used to regularize the infrared divergences (our results are 
$\lambda$ independent for small values of $\lambda$). The NLO Regge gluon trajectory (which defines the propagators of $t$-channel gluons) is 
\begin{eqnarray}
\label{eq:omega0}
\omega \left({\bf q}^2,\lambda^2 \right) &=& - \xi_\Theta \ln{\frac{{\bf q}^2}{\lambda^2}} + a^2 \frac{3}{2} \zeta (3).
\end{eqnarray}
The real emission part of the kernel is rather complicated~\cite{LevTolya} and reads
\begin{eqnarray}
\label{non_ang_av}
{\mathcal{K}} \left({\bf q},{\bf q}'\right) &=& 
\frac{a^2}{4 \pi} 
\left\{ -\frac{1}{({\bf q}-{\bf q'})^2}\ln^2{\frac{{\bf q}^2}{{\bf q'}^2}} \right. \nonumber\\
&&\hspace{-2cm}+\frac{2({\bf q}^2-{\bf q'}^2)}{({\bf q}-{\bf q'})^2({\bf q}+{\bf q'})^2} 
\left(\frac{1}{2}\ln{\frac{{\bf q}^2}{{\bf q'}^2}}
\ln{\frac{{\bf q}^2 {\bf q'}^2 ({\bf q}-{\bf q'})^4}
{({\bf q}^2+{\bf q'}^2)^4}} + \left( \int_0^{- \frac{{\bf q}^2}{{\bf q'}^2}} -
\int_0^{- \frac{{\bf q'}^2}{{\bf q}^2}} \right) 
dt \frac{\ln(1-t)}{t}\right)  \nonumber\\
&&\hspace{-2cm} \left. -\left(1-\frac{({\bf q}^2-{\bf q'}^2)^2}{({\bf q}-{\bf q'})^2
({\bf q}+{\bf q'})^2}\right) \left( \left( \int_0^1 
-\int_1^\infty \right) dz \frac{1}{({\bf q'}-z {\bf q})^2}
\ln{\frac{(z {\bf q})^2}{{\bf q'}^2}}\right) \right\}\nonumber\\
&&\hspace{-2cm}+\Theta \frac{a^2}{4 \pi} \left\{\frac{\left(3({\bf q}\cdot{\bf q'})^2-2 {\bf q}^2 {\bf q'}^2 \right)}{16 {\bf q}^2 {\bf q'}^2}\left(\frac{2}{{\bf q}^2}+\frac{2}{{\bf q'}^2}+\left(\frac{1}{{\bf q'}^2}-\frac{1}{{\bf q}^2}\right)\ln{\frac{{\bf q}^2}{{\bf q'}^2}}\right) \right. \nonumber\\
&&\hspace{-3cm} \left.  - \left(4-\frac{({\bf q}^2+{\bf q'}^2)^2}{8 {\bf q}^2 {\bf q'}^2}
- \frac{(2 {\bf q}^2 {\bf q'}^2 - 3 {\bf q}^4 - 3 {\bf q'}^4)}
{16 {\bf q}^4 {\bf q'}^4}({\bf q} \cdot {\bf q'})^2\right) 
\int^\infty_0 \frac{dx }{{\bf q}^2 + x^2 {\bf q'}^2} \ln{\left|\frac{1+x}{1-x}\right|}\right\}.
\end{eqnarray}

With all of this information it is now possible to use an iterative method and go back to rapidity space 
to obtain the following expression for the gluon Green function
\begin{eqnarray}
\label{ours}
f({\bf k}_a ,{\bf k}_b, Y) 
&=& \exp{\left(\omega \left({\bf k}_a^2,{\lambda^2}\right) Y \right)}
\left\{\frac{}{}\delta^{(2)} ({\bf k}_a - {\bf k}_b) \right. \\
&&\hspace{-2cm}+ \sum_{n=1}^{\infty} \prod_{i=1}^{n} 
\int d^2 {\bf k}_i \left[\frac{\theta\left({\bf k}_i^2 - \lambda^2\right)}{\pi {\bf k}_i^2} \xi  
+ {\mathcal{K}} \left({\bf k}_a+\sum_{l=0}^{i-1}{\bf k}_l,
{\bf k}_a+\sum_{l=1}^{i}{\bf k}_l\right)\frac{}{}\right]\nonumber\\
&& \hspace{-1cm} \times  
\int_0^{y_{i-1}} d y_i ~ {\rm exp}\left[\left(
\omega \left(\left({\bf k}_a+\sum_{l=1}^i {\bf k}_l\right)^2,\lambda^2
\right)\right.\right.\nonumber\\
&&\left.\left.\left.\hspace{0cm}
-\omega \left(\left({\bf k}_a+\sum_{l=1}^{i-1} {\bf k}_l\right)^2,
{\lambda^2}\right)\right) y_i\right] \delta^{(2)} \left(\sum_{l=1}^{n}{\bf k}_l 
+ {\bf k}_a - {\bf k}_b \right)\right\}, \nonumber
\end{eqnarray}
where $y_0\equiv Y$. We have obtained numerical results for this formula by performing a 
Monte Carlo integration of each of the terms in the sum, which implies to solve a large amount of 
nested integrals in transverse momentum and rapidity space. This is a rather complicated procedure 
where the correct sampling of the integrands plays a very important role, but which allows for a complete 
handling of the exclusive information in the parton ladder since we know the statistical weight of the 
different final state configurations. 

\begin{figure}[tbp]
  \centering
\epsfig{width=9cm,angle=0,file=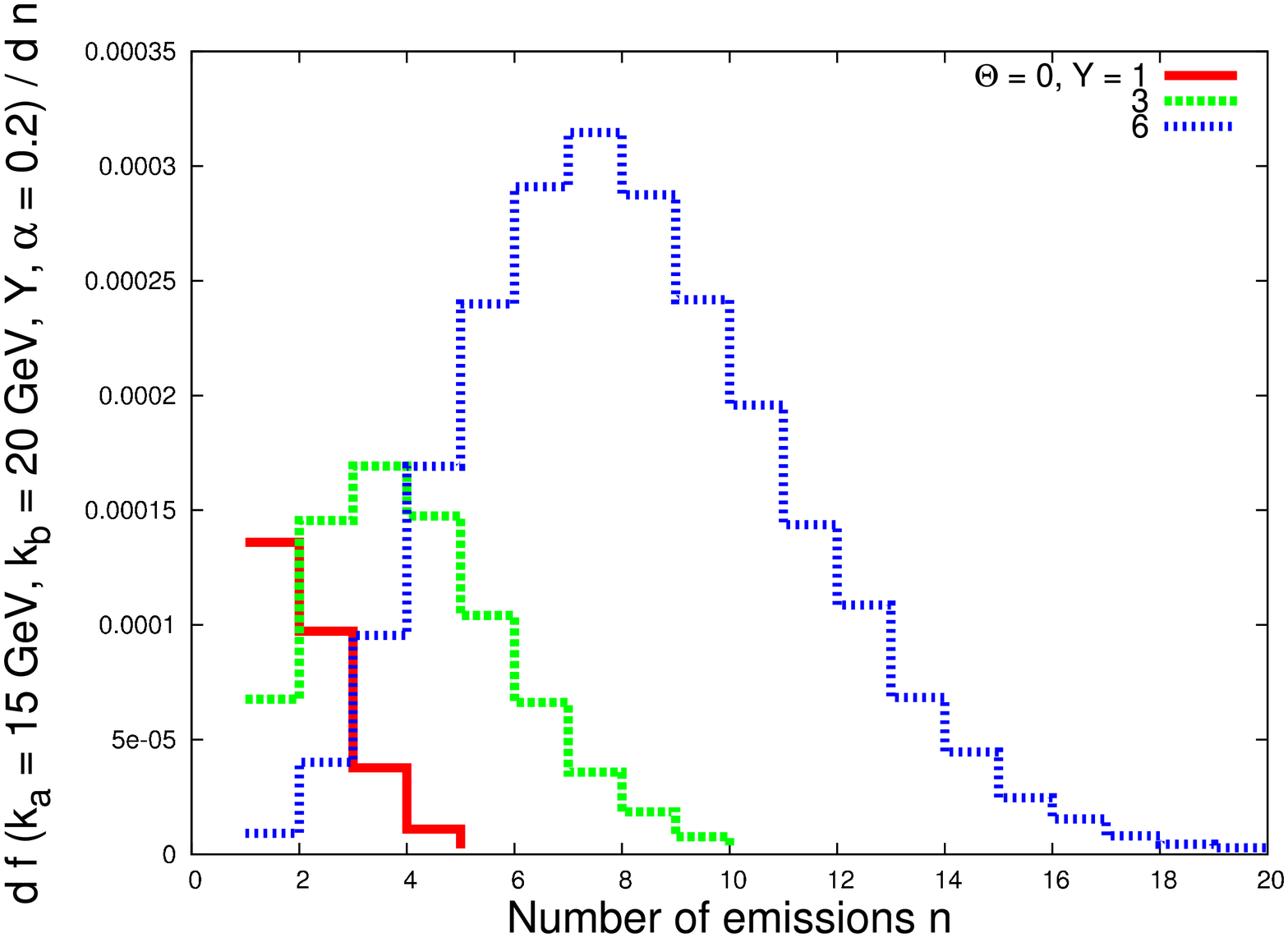}\\ 
\vspace{.3cm}
  \epsfig{width=9cm,angle=0,file=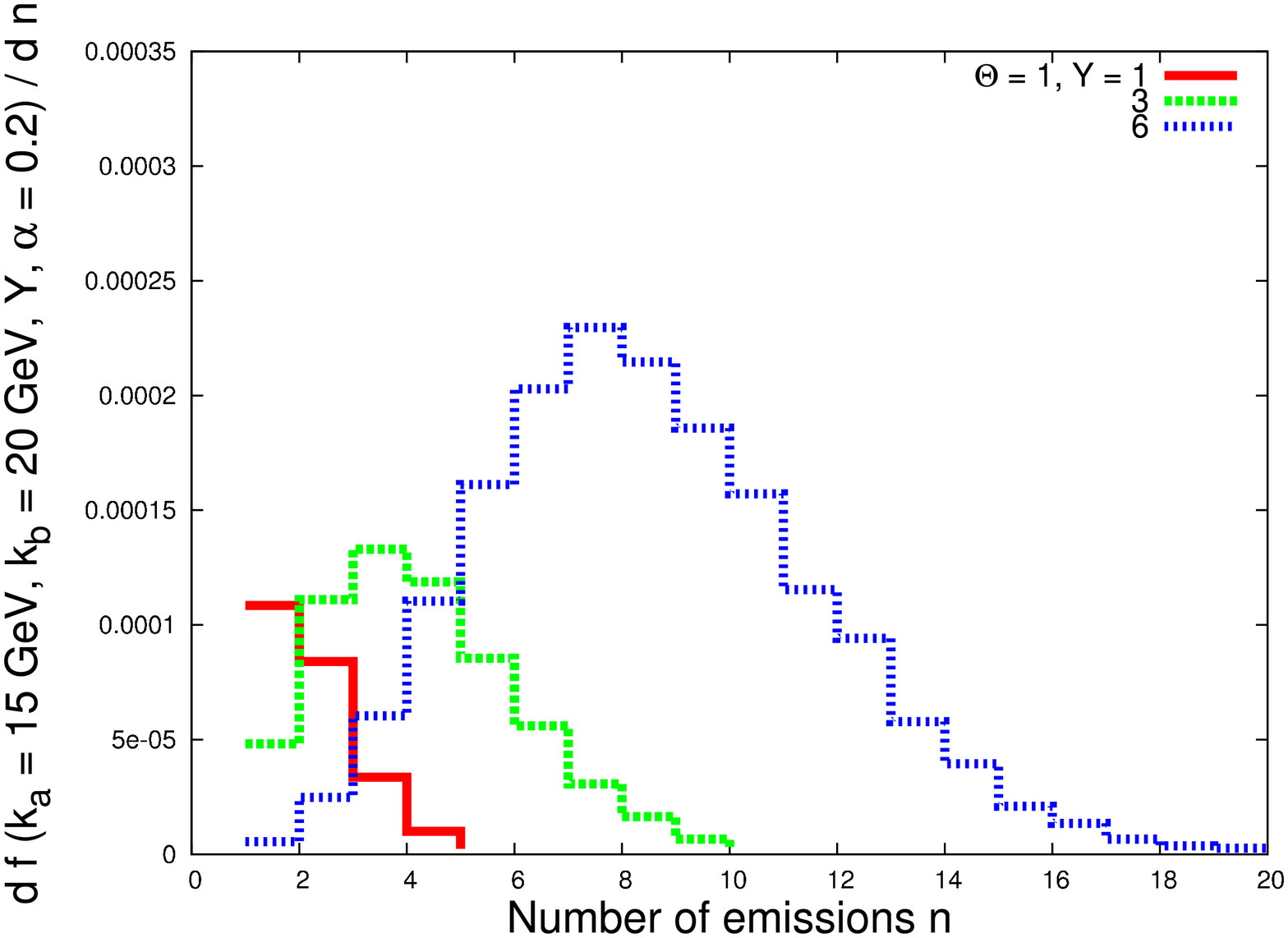}\\ 
\vspace{.3cm}
  \epsfig{width=9cm,angle=0,file=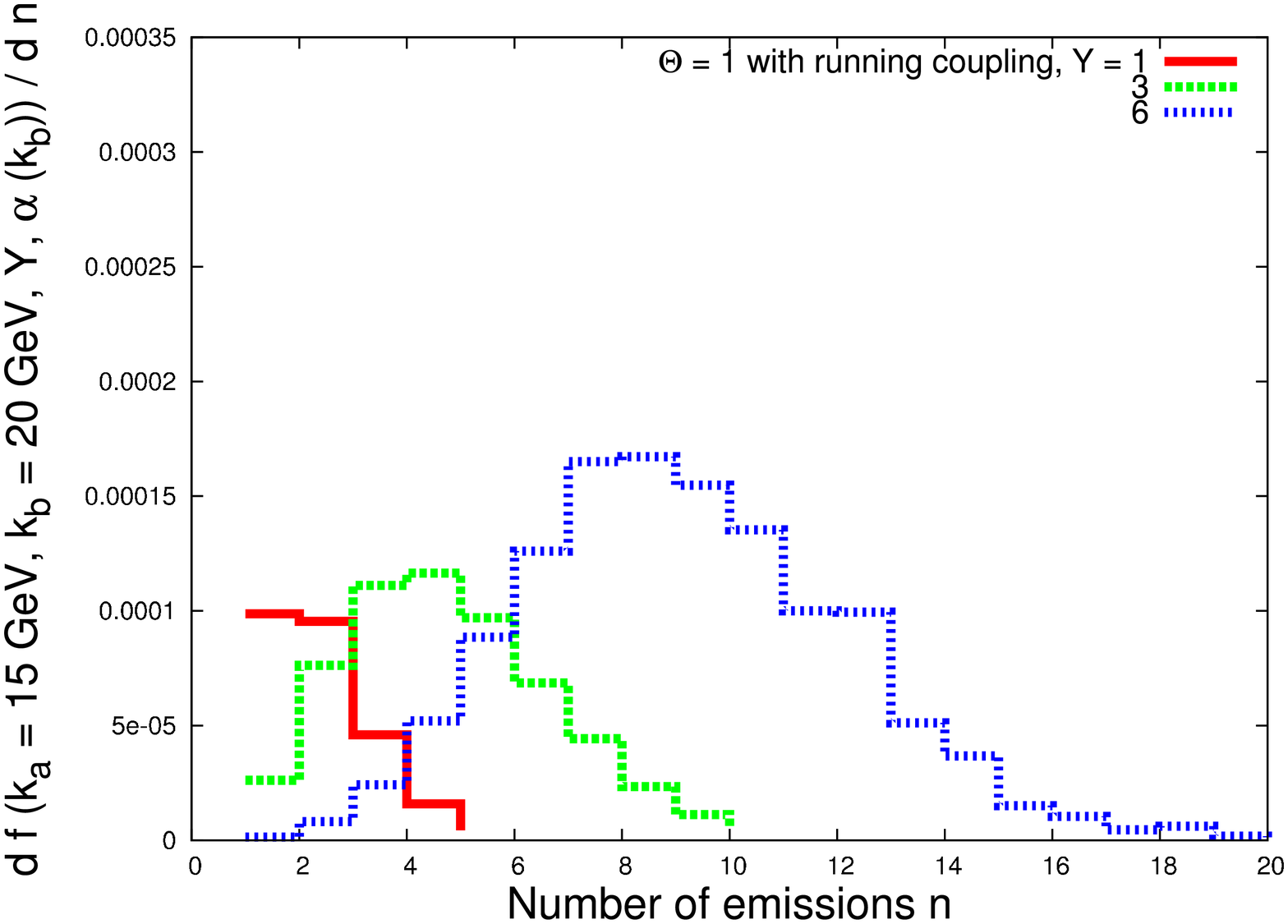}
  \caption{Multiplicity distribution in the number of emissions contributing to the gluon Green function. }
  \label{SUSYDiff-Emissions}
\end{figure}
We can check that the distribution in the number of iterations of the kernel needed to construct the gluon 
Green function does not vary, qualitatively, when scalars and gluinos are added to the gluon terms, 
which drive the multiplicity distribution (see Fig.~\ref{SUSYDiff-Emissions}, where the Green function 
corresponds to the area under the plots.). This statement is independent from introducing a running of the 
coupling in the gluon (QCD with no quarks) kernel (see last plot in Fig.~\ref{SUSYDiff-Emissions}).

\begin{figure}[tbp]
  \centering
  \epsfig{width=8.6cm,angle=-90,file=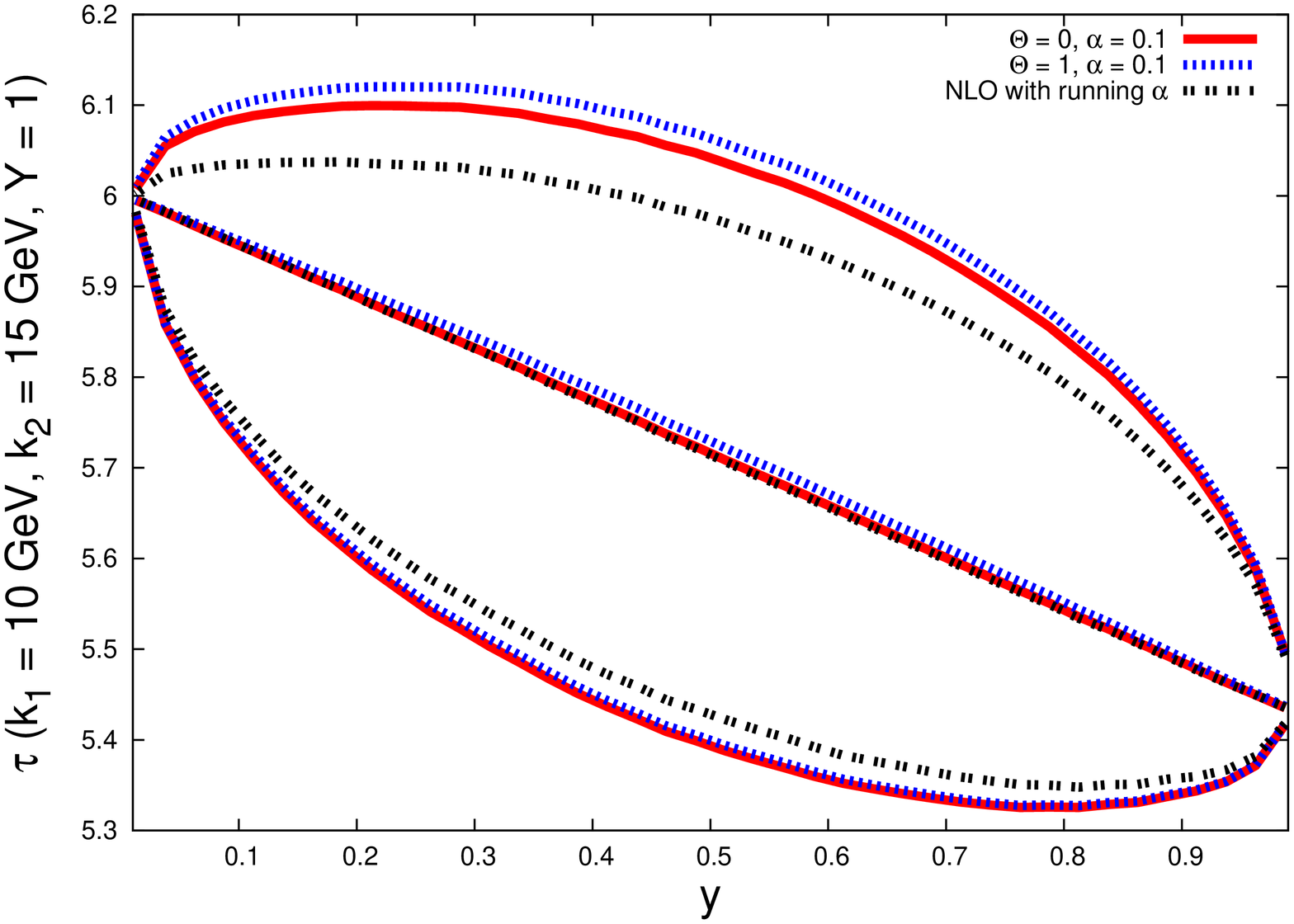}\\
\vspace{.5cm}
  \epsfig{width=12cm,angle=0,file=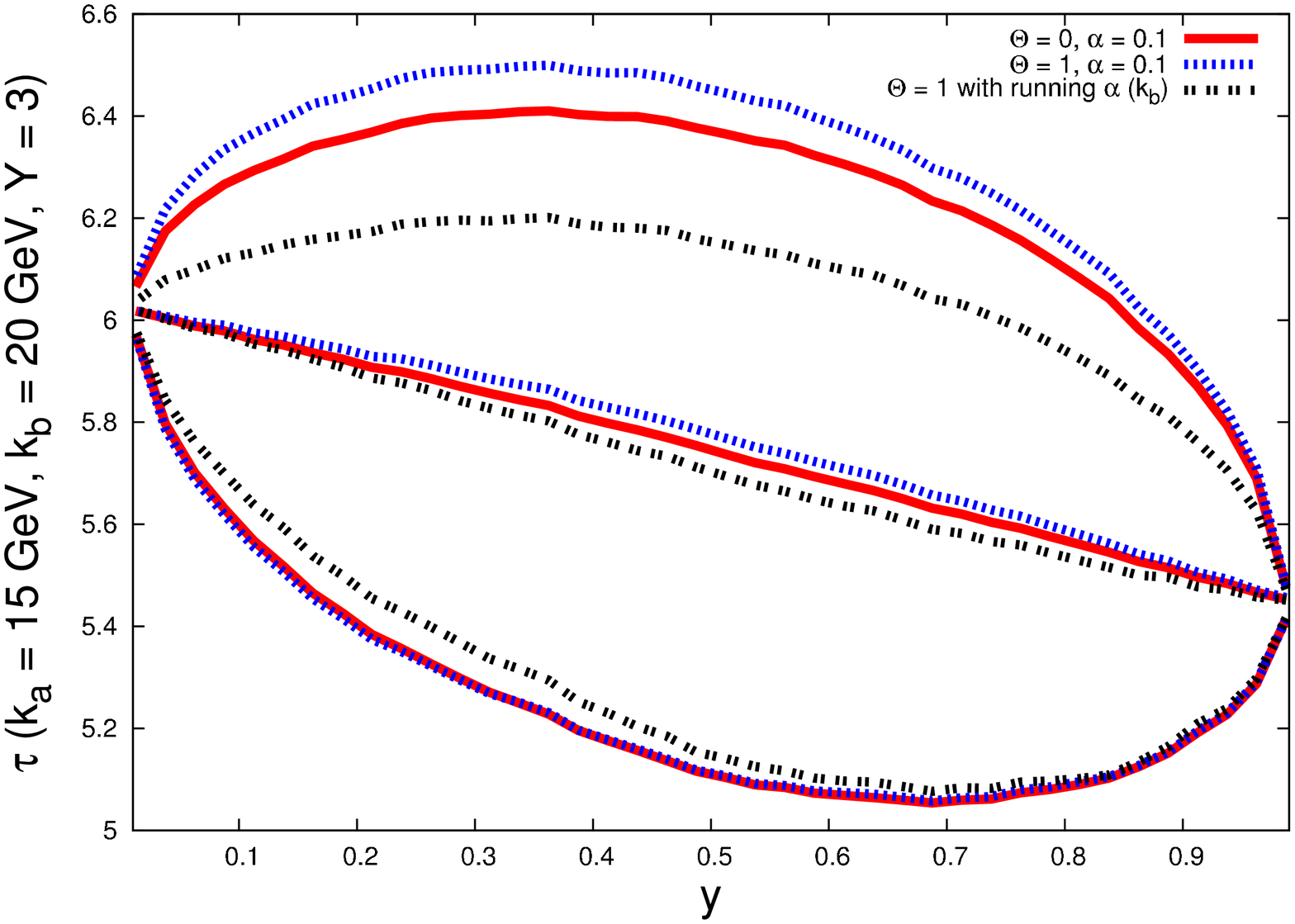}
  \caption{Diffusion plots for the propagation of internal modes into infrared and ultraviolet regions in 
  transverse momenta when constructing the gluon Green function. The upper plot corresponds to $Y=1$ and the lower one to $Y=3$.}
  \label{SUSYDiff-MelonY}
\end{figure}
It is also possible to find out what is the typical transverse momentum scale running in the internal propagators 
of the BFKL ladder. This is conveniently shown in Fig.~\ref{SUSYDiff-MelonY} where the mean value of the variable $\tau = \log{<p_i > / ({\rm GeV}^2)}$ is calculated (together with the lines of one standard deviation towards the infrared and ultraviolet) as a function of the normalized rapidities of the corresponding emitted 
particles. The main lesson to be taken from these plots is that the region with diffusion in the infrared is fully governed by the gluon dynamics in the SUSY kernel (setting $\Theta=0,1$ does not modify the lower lines) 
while in the ultraviolet region the scalars and gluinos do squeeze the plot downwards, decreasing the diffusion probability towards large scales.

\begin{flushleft}
{\bf \large Conclusions}
\end{flushleft}

In this work we have presented a study of the solution to the NLO BFKL equation in the $N=4$ supersymmetric Yang-Mills theory. The target has been to separate the ``QCD-like" gluon contributions 
from those stemming from scalar particles and gluinos. Our investigation has been performed at the level 
of the gluon Green function for the scattering of two off-shell reggeized gluons in quasi-multi-Regge kinematics. We have shown the good convergence in collinear regions of the perturbative series for higher Fourier components in the azimuthal angle (non-zero conformal spin $n$) and proven that the effect of the 
scalar and gluino diagrams is minimal for $n>0$ and very important for $n=0$, making the growth with energy of the forward amplitude to be much faster than for the purely gluon contributions. Using Monte Carlo integration techniques we showed that the scalars and gluinos do not affect the diffusion into 
the infrared in the so-called ``Bartels' cigar''  but force the average transverse momenta in the BFKL ladder to lie at less perturbative scales when looking at the ultraviolet diffusion sector. In future works it would be interesting to investigate theories with a lower number of supersymmetries, the 
NNLO version of the BFKL kernel and perform a similar study for the BFKL Pomeron at strong coupling. 

\noindent {\bf Acknowledgements}

F.C. thanks the Instituto de F\'isica Te\'orica UAM/CSIC for the warm hospitality. We acknowledge partial support from the European Comission under contract LHCPhenoNet (PITN-GA-2010-264564), the Comunidad de Madrid through HEPHACOS S2009/ESP-1473, and MICINN (FPA2010-17747) and Spanish MINECOs Centro de Excelencia Severo Ochoa Programme under grant SEV- 2012-0249. The work of F.C. was supported by European Commission, European Social Fund and Calabria Region, that disclaim any liability for the use that can be done of the information provided in this paper. G.C. thanks the support from the Research Executive 
Agency (REA) of the European Union under the Grant Agreement number PIEF-GA-2011-298582 and by 
MICINN (FPA2011-23778, FPA2007-60323 and CSD2007-00042 CPAN).

\end{document}